\documentclass[12pt]{article}

\newenvironment{pf}{\noindent\textbf{Proof.}\quad}{\hfill{$\Box$}}
\usepackage[colorlinks=false]{hyperref}
\usepackage[dvips]{graphicx}
\usepackage[usenames,dvipsnames]{color}
\usepackage{epsfig}
\usepackage{rotating}
\usepackage{amssymb}
\usepackage{amsmath,amsfonts}
\usepackage[latin1]{inputenc}
\usepackage{verbatim}
\usepackage[tableposition=top,font=small,labelfont=bf,format=hang]{caption}
\usepackage{booktabs}
\newtheorem{lem}{Lemma}
\newtheorem{thm}{Theorem}
\newtheorem{prop}{Proposition}

\newtheorem{cor}{Corollary}
\newcommand{\F}{\mathbb{F}}

\newcommand{\qed}{\hfill $\Box$ \\}

\begin{document}

\title{Constructions of optimal LCD codes over large finite fields}
\author{
Lin Sok\thanks{Lin Sok, School of Mathematical Sciences, Anhui University, Hefei, Anhui, 230601 and Department of Mathematics, Royal University of Phnom Penh, Cambodia, { \tt sok.lin@rupp.edu.kh}},
Minjia Shi\thanks{ Minjia Shi, Key Laboratory of Intelligent Computing Signal Processing, Ministry of Education, Anhui University, No.3 Feixi Road, Hefei, Anhui, 230039, China, School of Mathematical Sciences, Anhui University, Hefei, Anhui, 230601,
China and National Mobile Communications Research Laboratory, Southeast University, China, {\tt smjwcl.good@163.com}}, and
Patrick Sol\'e\thanks{CNRS/LAGA, University of Paris 8, 93 526 Saint-Denis, France , {\tt sole@enst.fr}}
}
\date{}
\maketitle
\begin{abstract}
In this paper, we prove existence of optimal complementary dual codes (LCD codes) over large finite fields. 
We also  give methods to generate orthogonal matrices over finite fields and then apply them to construct LCD codes.
Construction methods include random sampling in the orthogonal group, code extension, matrix product codes and projection over a self-dual basis.

\end{abstract}
\section{Introduction}
Linear codes with complementary duals, which
we refer to as LCD codes, were introduced by Massey in
\cite{Massey}. They give an optimum linear coding solution for the two
user binary adder channel. They are also used in counter measures to
passive and active side channel analyses on embedded crypto-systems,
see \cite{CarGui} for a detailed description. It is known from \cite{S} that LCD codes are asymptotically good.

Dougherty et al. \cite{DKOSS} constructed binary LCD codes using orthogonal matrices, self-dual codes, combinatorial designs and Gray map from codes over a family of non chain rings
of characteristic $2$. Liu et al. \cite{LL} characterized matrix product linear complementary dual (MPLCD) codes and gave their constructions from orthogonal-like matrices.
Using generalized Reed-Solomon codes, the authors of \cite{CheLiu} and \cite{Jin} proved the existence of optimal LCD codes over finite fields with some conditions on lengths and the field sizes. The problem of existence of $q-$ary $[n,k]$ MDS LCD codes has completely been solved by Carlet et al. \cite{CMCT} for the Euclidean case.

Recently, in the paper \cite{SSS}, MDS self-dual codes over large prime fields have been constructed from orthogonal matrices and from the generalized method of \cite{AG}. It is important 
to note that a single orthogonal matrix gives rise to several LCD codes, by  a different choice of basis.

From the existence of MDS self-dual codes for example in \cite{GraGul,JinXin} as well as from MDS self-orthgonal codes, we construct MDS LCD codes with certain lengths. We also generalize
the constructions \cite{SSS} of orthogonal matrices from prime fields to arbitrary finite fields and afterwards we give explicit constructions of LCD and MPLCD codes. Short LCD codes are constructed from orthogonal-like matrices by randomly sampling elements in the orthogonal group and from code extension by two symbols while the long ones are constructed from matrix product codes and from projection over a self-dual basis. Many optimal, almost MDS and MDS codes over different fields are obtained. 

The paper is organized as follows: Section II gives preliminaries for LCD codes. Section III proves the existence of LCD codes of certain lengths and also gives method to construct and to extend  an LCD code. In Section IV we present numerical results of some optimal codes, almost MDS and MDS LCD codes over different large fields.

\section{Preliminaries}
\label{Sec-Prelim}
 A {\em
linear $[n,k]$ code $C$ of length $n$} over ${{\mathbb F}_q}$ is a $k$-dimensional subspace
of  $ {\mathbb F}_q^n$. An element in $C$ is called a {\em codeword}. The
(Hamming) weight wt$({\bf{x}})$ of a vector ${\bf{x}}=(x_1, \dots,
x_n)$ is the number of non-zero coordinates in it. The {\em minimum
distance ({\rm{or}} minimum weight) $d(C)$} of $C$ is
$d(C):=\min\{{\mbox{wt}}({\bf{x}})~|~ {\bf{x}} \in C, {\bf{x}} \ne
{\bf{0}} \}$. The {\em Euclidean inner product} of ${\bf{x}}=(x_1,
\dots, x_n)$ and ${\bf{y}}=(y_1, \dots, y_n)$ in ${\mathbb F}_q^n$ is
${\bf{x}}\cdot{\bf{y}}=\sum_{i=1}^n x_i y_i$. The {\em dual} of $C$,
denoted by $C^{\perp}$ is the set of vectors orthogonal to every
codeword of $C$ under the Euclidean inner product. A linear code
$C$ is called linear complementary dual ({\em LCD}) if $C\cap C^{\perp}=\{0\}$. One among the important parameters for a code is its minimum distance. If $C$ is a linear $[n,k]_{{\mathbb F}_q}$code, then from the Singleton bound, its minimum distance is bounded by
$$d(C)\le n-k+1.$$
A linear code meeting the above bound is called {\em Maximum Distance Separable} ({MDS}) code.  A linear $[n,k]$code $C$ is called {\em almost} MDS if $d(C)=n-k$. A code is called {\it optimal} if it has the highest possible minimum distance for its length and dimension and thus an MDS code is optimal. The following result is due to MacWilliams and Sloane \cite{MacSlo}.
\begin{thm}[\cite{MacSlo}]
Let $C$ be an $[n, k, d]$ code over $\F_q.$ The following statements are equivalent:
\begin{enumerate} 
\item $C$ is $MDS;$
\item $C^\perp$ is $MDS.$
\end{enumerate}
\end{thm}
\section{Construction of LCD codes}
The following lemmas characterize LCD codes.
 \begin{lem}[\cite{DKOSS}]\label{lem1}
 Let ${\bf u}_1,{\bf u}_2,\dots,{\bf u}_k$ be vectors over a commutative ring $R$ such that ${\bf u}_i.{\bf u}_i=1$  for each $i$ and ${\bf u}_i.{\bf u}_j=0$ for $i \neq j.$ Then
    $C = \langle  {\bf u}_1,{\bf u}_2,\dots,{\bf u}_k \rangle $ is an LCD code over $R$.
    \end{lem}
    \begin{lem}[\cite{Massey}]\label{lem2}
    Let $G$ be a generator matrix for a code over a field.  Then $det(GG^{\top}) \neq 0$ if and only if $G$ generates an LCD code.
    \end{lem}
{\bf Remark} Lemma \ref{lem1} and Lemma \ref{lem2} show that if an $n\times n$ matrix $A$ is orthogonal or orthogonal-like, i.e, $AA^{\top}$ is a diagonal matrix with all diagonal entries nonzero, then any row(s) of $A$ generate an LCD code. 
\begin{lem}\label{lem3} Assume that there exists an MDS self-orthogonal $[n,k]$ code over $\F_q$. Then there exists an MDS LCD $[n-k,k']$ code for $1\le k'\le k$.\end{lem}
{\pf Assume that there exists a self-orthogonal code $C$ with parameter $[n,k,n-k+1]$ and let $G$ be its generator matrix in the systematic form. $$G=(I_k|A).$$
Take $C_0^{(k')}$ as a subcode of $C$ generated by the last $k'$ rows of $G$. So $C_0^{(k')}$ is an $[n,k',\ge n-k+1]$ code. Shortening the first $k-k'$ positions of $C_0^{(k')}$, we get a code $C_{00}^{(k')}=[n-k+k',k', n-k+1]$ with its generator matrix $$G_{00}=(I_{k'}|A^{(k')}),$$
where $A^{(k')}$ is the submatrix of $A$ with the last $k'$ rows. Now puncturing the first $k'$ positions of $C_{00}^{(k')}$, we get, by the Singleton bound, a code $C'_{000}=[n-k,k', n-k+1-k' ]$ whose generator matrix is of the form
$$G_{000}=A^{(k')}.$$
Finally the result follows from Lemma \ref{lem1}. \qed
}

\begin{thm} We have the following existence results:
\begin{enumerate} 
\item For any even prime power $q=2^m$, there exists an MDS LCD $[n,k]$ code for $1\le n\le 2^{m-1},1\le k \le n$.\\
\item For any odd prime power $q$ there exists an MDS LCD $[n,k]$ code, for $1\le k \le n,$ with the following conditions.
\begin{enumerate}
\item  $n=(q+1)/2$,
\item $q\equiv 1 \pmod 4$ $q\ge 2^{(2n)} \times (2n)^2$,
\item  $q=r^2$ and $2n\le r$,
\item $ q=r^2$ and $2n-1$ is an odd divisor of $q-1,$
\item  $r\equiv 3 \pmod 4$ and $n=tr$ for any $t\le (q-1)/2.$
\end{enumerate}
\end{enumerate}
\end{thm}
{\pf $1.$ and $2.(a)$ follow from \cite{GraGul} and Lemma \ref{lem3}, where as $2.(b)-2.(e)$ follow from \cite{JinXin} and Lemma \ref{lem3}.}
\begin{thm} Let $q=p^m, m> 1$ for some prime $p$, $n|q-1$ and $k\le \lfloor (n-1)/2 \rfloor$. Then there exists an MDS LCD $[n-k,k']$ code for $1\le k' \le k.$
\end{thm}
{\pf Let $q=p^m, m> 1$ for some prime $p$ and $n|q-1$. Let $\alpha$ be a primitive $n-$th root of unity. From the BCH bound, the polynomial $g(x)= \prod\limits_{i=1}^{k} (x-\alpha^i)$ generates an MDS code $D=[n,n-k]$ over $\F_q$ with its defining set $T_D=\{1,\hdots,k\}.$ It implies, from [\cite{HufPle} 
Theorem 4.4.9 ], that the dual of $D$ is an MDS code $C=[n,k]$ with its defining set $T_C=\{1,\hdots,n-k-1\}\cup \{0\}$. Since $k\le \lfloor (n-1)/2 \rfloor$, $C$ is self-orthogonal. Finally the result follows from Lemma \ref{lem3}. \qed
}

The rest of our constructions are based on orthogonal matrices. In what follows, we present some elements used to generate an orthogonal group. In the sequel, $\mathbb{F}_q$ denotes a finite field of characteristic $p$, that is $q=p^m$ for some positive integer $m$.

The {\em orthogonal group} of index $n$ over a finite field with $q$ elements is defined by
$${\cal O}_n(q):=\{A\in GL(n,q)| AA^{\top}=I_n\}.$$
Let ${\cal P}_n$ be the set of $n\times n$ permutation matrices. 

For $q=2$, with the convention ${\cal O}_n:={\cal O}_n(2),$ we have the following theorem due to Janusz \cite{J}.
\begin{thm}[\cite{J}]{\label{jan.2}}
The orthogonal groups ${\cal O}_n$  are generated as follows
\begin{enumerate}
\item for $1\leq n\leq 3$, ${\cal O}_n={\cal P}_n$,
\item for $n \geq 4$, ${\cal O}_n=\langle {\cal P}_n , {T}_{\bf u}\rangle, $
\end{enumerate}
where $\bf u$ is a binary vector of Hamming weight $4$ and $T_{\bf u}$ is the transvection defined by
$$\begin{array}{ll}
T_{{\bf u}}: &\mathbb{F}_2^n \longrightarrow \mathbb{F}_2^n\\
 &{\bf x} \mapsto
({\bf x.u}){\bf u}.
\end{array}
$$
\end{thm}
To generate elements of orthogonal group in any arbitrary finite field, we give more general setting as follows. 
Let $q=p^{m}$ for some prime $p$ and some positive integer $m.$ Let $\theta=\frac{p-1}{2}\in {\mathbb F}_p$ if $p\neq 2$ and $\theta=1$ otherwise. Let $\alpha,\beta\in {\mathbb F}_q\backslash \{0\}$ such that $\alpha^2+\beta^2=1$ and ${\bf v}=(\alpha-1){\bf e}_1+\beta {\bf e}_2,{\bf w}=-\beta{\bf e}_1+(\alpha-1) {\bf e}_2$. Let ${\bf u}={\bf e}_1+{\bf e}_2+{\bf e}_3+{\bf e}_4$ if $n\ge 4$, where $\{ {\bf e}_1,\hdots,{\bf e}_n \}$ is the canonical basis of $\mathbb{F}_q^n$. Define two linear maps
$$\begin{array}{llll}
T_{{\bf u},\theta}: &\mathbb{F}_q^n \longrightarrow \mathbb{F}_q^n,& T_{{\alpha},{\beta}}:& \mathbb{F}_q^n \longrightarrow \mathbb{F}_q^n\\
 &{\bf x} \mapsto \theta({\bf x.u}){\bf u}& &{\bf x} \mapsto {\bf x}+({\bf x.{\bf v}}){\bf e}_1+({\bf x}.{\bf w}){\bf e}_2.\\
\end{array}
$$
Denote
$${\cal T}_n{(q)}:=
\begin{cases}
\langle {\cal P}_n ,T_{{\alpha},{\beta}}\rangle\text{ if } n\leq 3,\\
\langle {\cal P}_n ,T_{{\alpha},{\beta}},T_{{\bf u},\theta}\rangle, \text{ otherwise}.
\end{cases}$$
It is well-known that transvections are linear and of order $2$ and thus ${\cal T}_n{(q)}$ is a subgroup of ${\cal O}_n(q).$ The orders of 
${\cal T}_n{(q)}$ are calculated in \cite{mag} and compared with \cite{Mac} in Table \ref{Table:1} for some values $3\leq q\leq 25$ and $n=4,5$.\\

\noindent
{\bf Conjecture} For any $q$ and $n\ge 5,{\cal O}_{n}{(q)}= {\cal T}_{n}{(q)}.$\\

\begin{table}\centering{\caption{Comparison between the orders $|{\cal T}_n{(q)}|$ and $|{\cal O}_n(q)|$ for $3\le q \le 25$} and $n=4,5$ \label{Table:1}}
$$
\begin{array}{|c|c|c|c|c|}
\hline
q&|{\cal T}_4{(q)}|\cite{mag}&|{\cal O}_4{(q)}|\cite{Mac}&|{\cal T}_5{(q)}|\cite{mag}&|{\cal O}_5{(q)}|\cite{Mac}\\
\hline
3&384&
1152&
103680&
103680\\
4&3840&
3840&
979200&
979200\\
5&384&
28800&
18720000&
18720000\\
7&225792&
225792&
553190400&
553190400\\
8&258048&
258048&
1056706560&
1056706560\\
9&1036800&
1036800&
6886425600&
6886425600\\
11&3484800&
3484800&
51442617600&
51442617600\\
13&9539712&
9539712&
274075925760&
274075925760\\
16&16711680&
16711680&
1095199948800&
1095199948800\\
17&47941632&
47941632&
4017988177920&
4017988177920\\
19&93571200&
93571200&12228071558400&12228071558400\\
23&294953472&
294953472&82966104944640&82966104944640\\
25&486720000&
486720000&190429200000000&190429200000000\\
\hline
\end{array}
$$
\end{table}


Now we introduce some constructions of LCD  and matrix product LCD codes from orthogonal matrices. 
\begin{prop} Let $A\in {\cal O}_n(q)$ and $A_k$ a submatrix obtained from $A$ by keeping  $k$ rows. Then
the matrix 
\begin{equation}\label{eq:01}
G=A_k
\end{equation}
generates an LCD code.
\end{prop}
{ \pf Since $A$ is orthogonal, the result follows from Lemma \ref{lem1}. \qed}
\begin{prop} Let $A\in {\cal O}_n(q)$ and $A_k$ a submatrix obtained from $A$ by keeping  $k$ rows. Then for any  $\lambda_1,\hdots,\lambda_k\in {\mathbb F}_q\backslash \{0\}$,
the matrix 
\begin{equation}\label{eq:02}
G=diag(\lambda_1,\hdots,\lambda_k)A_k
\end{equation}
generates an LCD code.
\end{prop}
{ \pf Since $A$ is orthogonal, $GG^{\top}$ is orthogonal-like and thus the result follows from Lemma \ref{lem2}. \qed}
\begin{prop} Let $A\in {\cal O}_n(q)$ and $A_k$ a submatrix obtained from $A$ by keeping  $k$ rows, with $k$ being even. Let $(\alpha_i)_{1\le i \le k/2},(\beta_i)_{1\le i \le k/2}\in {\mathbb F}_q\backslash \{0\}$ such that $\alpha_i^2+\beta_i^2\neq 0$ and $D_i=\left(
\begin{array}{cc}
 \alpha_i&\beta_i\\
- \beta_i&\alpha_i
\end{array}
 \right)$. Then for any $\lambda_1,\hdots,\lambda_k\in {\mathbb F}_q\backslash \{0\}$
the matrix 
\begin{equation}\label{eq:03}
G=diag(\lambda_1,\hdots,\lambda_k)diag(D_1,\hdots,D_{k/2})A_k
\end{equation}
 generates an LCD code.\\
Moreover if $k$ is odd then the matrix 
\begin{equation}\label{eq:03}
G'=diag(\lambda_1,\hdots,\lambda_k)diag(D_1,\hdots,D_{(k-1)/2},1)A_k
\end{equation}
also generates an LCD code.
\end{prop}
{ \pf Since $A$ is orthogonal, $GG^{\top}$ is orthogonal-like and thus the result follows from Lemma \ref{lem2}. \qed}
{\bf Remark}
In the above constructions, for $n$ large, in practice the orthogonal matrix $A$ are randomly sampled from ${\cal O}_n(q)$.

In what follows, we construct longer LCD codes by coordinate extension of shorter codes whose generator matrices are rows of orthogonal matrices. Note that if $C_n$ is a linear $[n,k,d]$ code then $C_n$ can be decomposed as a direct sum
$C_n=D \oplus E$, where $D$ (resp. $E$) is a subcode of $C_n$ of minimum weight $d$ (resp. $e>d$). Moreover the generator matrix $G_n$ of $C_n$ can be written as:
\begin{equation}\label{eq:1}
G_n=\left(
\begin{array}{c}
G_d\\
G_e\\
\end{array}
\right).
\end{equation}
This decomposition allows us to reduce the complexity of searching good codes efficiently when we want to construct LCD $[n,k,\ge d]$ codes from an LCD $[n,k,d]$ code 
and when the dimension of the subcode $D$ is large. For example assume that  there exist scalars $a,b\in {\mathbb F}_q\backslash\{0\}$ such that $a^2+b^2\equiv 0 \pmod q$.
From this data, an LCD code of length $n+2$ can be obtained, by extending two coordinates, from an LCD code of length $n$ with its generator matrix $G_n$ of the above form (\ref{eq:1}) as follows.
\begin{equation}
\left(\begin{array}{ccccc}
& & &\alpha_1a&\alpha_1b\\
& & &\alpha_2(-b)&\alpha_2a\\
& G_d&&\vdots&\vdots\\
&  & &\alpha_{2i-1}a&\alpha_{2i-1}b\\
&  & &\alpha_{2i}(-b)&\alpha_{2i}a\\
& &&\vdots&\vdots\\
\hline
& & &\beta_1a&\beta_1b\\
& & &\beta_2(-b)&\beta_2a\\
& G_e&&\vdots&\vdots\\
&  & &\beta_{2i-1}a&\beta_{2i-1}b\\
&  & &\beta_{2i}(-b)&\beta_{2i}a\\
& &&\vdots&\vdots\\
\end{array}\right),
\end{equation}
where $\alpha_1,\alpha_2, \hdots \in {\mathbb F}_q\backslash \{0\}$ and $\beta_1,\beta_2, \hdots \in {\mathbb F}_q.$ It should be noted that for $q$ being prime and $q\equiv 3 \pmod 4$ such $(a,b)\neq (0,0)$ does not exist, otherwise there exists $c\in {\mathbb F}_q $ with $c^2+1\equiv 0\pmod q.$

The following propositions construct LCD $[n+2,k]$ codes from an LCD $[n,k]$ code, which can later be completed by a direct sum with a one-dimensional code to produce LCD $[n+2,k+1]$ codes.
\begin{prop}   Let $C_n$ be an LCD code $[n,k,d]$ over ${\mathbb F}_q$ with its generator matrix $G_n$ being rows of an orthogonal matrix. Assume that there exist $a,b\in {\mathbb F}_q\backslash\{0\}$ such that $a^2+b^2\equiv 0 \pmod q$. Then for any $\lambda_1,\hdots, \lambda_n \in {\mathbb F}_q$, an extended code ${\bar{C}_n}$ of $C_n$ with the following generator matrix $G_{\bar{C}_n}$ is an LCD code $[n+2,k,\ge d]$:
\begin{equation} \label{eq:recursive1}
G_{\bar{C}_n}=\left(\begin{array}{ccccc}
& & &\lambda_1a&\lambda_1b\\
& & &\lambda_2(-b)&\lambda_2a\\
& G_n&&\vdots&\vdots\\
&  & &\lambda_{2i-1}a&\lambda_{2i-1}b\\
&  & &\lambda_{2i}(-b)&\lambda_{2i}a\\
& &&\vdots&\vdots\\
\end{array}\right).
\end{equation}
\end{prop}
{\pf Since $C_n$ is LCD,  with the assumption $a^2+b^2\equiv 0 \pmod q$, each row of $G_{\bar{C}_n}$, which is an extended row of $G_n$, is orthogonal to itself and to the other rows and thus the result follows.
\qed

Recall that the matrix-product code $C = [C_1,\hdots,C_l]A$ is defined as a linear code whose all codewords are matrix product $[c_1,\hdots,c_l]A,$ where $c_i \in C_i$ is an $n\times 1$ column vector and $A = (a_{ij})_{l\times m}$ is an $l\times m$ matrix over ${\mathbb F}_q.$ Here $l\leq  m$ and $C_i$ is an $[n,| C_i |]_{{\mathbb F}_q}$ code over ${\mathbb F}_q.$ If $C_1,\hdots,C_l$ are linear with generator matrices $G_1,\hdots,G_l,$ respectively, then $[C_1,\hdots,C_l]A$ is linear with generator matrix
$$
G=\left(
\begin{array}{cccc}
a_{11}G_1 &a_{12}G_1&\cdots&a_{1m}G_1\\
a_{21}G_2 &a_{22}G_2&\cdots&a_{2m}G_2\\
\vdots&\vdots&\cdots&\vdots\\
a_{l1}G_l &a_{l2}G_l&\cdots&a_{lm}G_l\\

\end{array}
\right).
$$
A matrix-product code $C=[C_1,\hdots,C_l]A$ over ${\mathbb F}_q$ which is linear complementary dual is called matrix-product linear complementary code (MPLCD) over ${\mathbb F}_q.$

The two following lemmas are vital for the construction of MPLCD codes.
\begin{lem} [\cite{LS}] Let  $(C_i)_{1\le i \le l}$ be linear codes over $F_q$ with parameters $[n, k_i]$  and $A$ be 
an $l\times m$ matrix of full row rank. Then $C=[C_1,\hdots, C_l]A$ is an $[mn,\sum_{i=1}^lk_i]$ code. 
\end{lem}
\begin{lem}[\cite{BN}]\label{lem:BN} Let  $(C_i)_{1\le i \le l}$ be linear codes over $F_q$ with parameters $[n, k_i]$  and $A$ be a non-singular matrix. If $C = [C_1,\hdots,C_l]A,$ then
$([C_1,\hdots,C_l]A)^\perp = [C_1^\perp,\hdots,C_l^\perp](A^{-1})^{\top}.$
\end{lem}
MPLCD codes can be now characterized as follows.
\begin{prop}\label{prop:MPLCD1} Let $C_1,C_2,\hdots ,C_l$ be linear codes over ${\mathbb F}_q$.  Let $A\in {\cal O}_l(q)$ and Then $C=[C_1,C_2,\hdots,C_l]A$ is an MPLCD code if and only if $C_1,C_2,\hdots,C_l$ are all LCD codes.
\end{prop}
{\pf From Lemma \ref{lem:BN} for any $A\in {\cal O}_l(q)$, $\left([C_1,\hdots,C_l]A\right)^\perp=[C_1^\perp,\hdots,C_l^\perp]A$.

Assume there exists $1\le i \le l$ such that $C_i$ is not LCD. Then both $C_i$ and $C_i^\perp$ contains 
a nonzero codeword $c_i$. Now $[0,\hdots,c_i,\hdots,0]A\in C\cap C^\perp$ with $C=[C_1,\hdots,C_i,\hdots,C_l]A$, which is a contradiction.

Conversely assume that $C=[C_1,\hdots,C_l]A$ is not MPLCD. Then there exists $(c_1,\hdots,c_l)\neq (0,\hdots,0)$ such that $[c_1,\hdots,c_l]A \in C\cap C^\perp $. Thus $C_i\cap C_i^\perp $ contains a nonzero codeword $c_i$ for some $1\le i \leq l$, which is again a contradiction.
\qed
}

Similarly we have the following characterization.
\begin{prop} \label{prop:MPLCD2}Let $C_1,C_2,\hdots ,C_l$ be linear codes over ${\mathbb F}_q$.  Let $A\in {\cal O}_l(q)$ and $\bar{A}=diag(a_1,\hdots,a_l)A$ with $a_1,\hdots,a_l \in {\mathbb F}_q\backslash \{0\}$. Then $C=[C_1,C_2,\hdots,C_l]\bar{A}$ is an MPLCD code if and only if $C_1,C_2,\hdots,C_l$ are all LCD codes.
\end{prop}
{ \pf Since $\bar{A}$ is  a non-singular matrix, from Lemma \ref{lem:BN}, $\left([C_1,\hdots,C_l]\bar{A}\right)^\perp=[C_1^\perp,\hdots,C_l^\perp]({\bar{A}}^{-1})^{\top}$. Now $({\bar{A}}^{-1})^{\top}=D\bar {A}$, where $D=diag(a_1^{-2},\hdots,a_l^{-2})$. Since $C_i^\perp=a_i^{-2}C_i^\perp$ for $1\le i \le l$, we get 
$$\left([C_1,\hdots,C_l]\bar{A}\right)^\perp=[C_1^\perp,\hdots,C_l^\perp]({\bar{A}}^{-1})^{\top}=[C_1^\perp,\hdots,C_l^\perp]D{\bar{A}}=[C_1^\perp,\hdots,C_l^\perp]{\bar{A}}.$$
Now the necessary and sufficient condition can be proved in the same way as in Proposition \ref{prop:MPLCD1}.
\qed}
A more generalized case is to consider other orthogonal-like matrices. Let $A\in {\cal O}_l(q)$ and $\lambda_1,\hdots,\lambda_l\in {\mathbb F}_q\backslash \{0\}$. Let $(\alpha_i)_{1\le i \le \lfloor l/2 \rfloor},(\beta_i)_{1\le i \le \lfloor l/2 \rfloor}\in {\mathbb F}_q\backslash \{0\}$ such that $\alpha_i^2+\beta_i^2\neq 0$ and $D_i=\left(
\begin{array}{cc}
 \alpha_i&\beta_i\\
- \beta_i&\alpha_i
\end{array}
 \right)$.
Set
\begin{equation}\label{eq:04}
\bar{A}=
\begin{cases}
diag(\lambda_1,\hdots,\lambda_l)diag(D_1,\hdots,D_{\lfloor l/2\rfloor })A,\text{ if } $l$ \text{ is even,}
\\
diag(\lambda_1,\hdots,\lambda_l)diag(D_1,\hdots,D_{\lfloor l/2\rfloor },1)A,\text{ otherwise.}
\end{cases}
\end{equation}

\begin{cor} Let $C_1,C_2,\hdots ,C_l$ be linear codes over ${\mathbb F}_q$.  Let $\bar{A}$ be as in (\ref{eq:04}). Then $C=[C_1,C_2,\hdots,C_l]\bar{A}$ is an MPLCD code 
if and only if $C_1,C_2,\hdots,C_l$ are all LCD codes.
\end{cor}

Let $B=\{e_0,e_1, \cdots, e_{\ell-1}\}$ be a self-dual basis of $\mathbb{F}_{q^\ell}$ over $\mathbb{F}_q$, that is, $$\rm{Tr}(e_i,e_{j})=\delta_{i,j},$$ where 
$\rm{Tr}$ denotes the trace of $\mathbb{F}_{q^\ell}$ down to $\mathbb{F}_q$ and $\delta_{i,j}$ is the Kronecker symbol. Define the map
$\phi_B:\mathbb{F}_{q^\ell}\longrightarrow \mathbb{F}_{q}^{\ell},$ by the formula $$ \sum_{i=0}^{\ell-1}a_ie_i\mapsto (a_0,\hdots,a_{\ell-1}),$$
and extend $\phi$ to $\mathbb{F}_{q^\ell}^n$ in the natural way.

\begin{thm} A linear code $C$ of length $n$ over $\mathbb{F}_{{q}^{\ell}}$ is LCD if and only if the linear code $\phi_B(C)$ of length $n\ell$ over $\mathbb{F}_{q}$ is LCD.
\end{thm}

{\pf See \cite{GuOzSo}.}
\section{Some numerical results}
To illustrate our constructions from orthogonal matrices, we present some numerical results of optmal codes obtained. The generator matrices are available from the author.

We list parameters of LCD codes obtained from our construction in Tables 2--5. Here we describe our constructions:\\

\noindent
$\bullet$ Constructions $(1)-(4)$ use random sampling. To obtain the optimal LCD codes with minimum distance $d$ from these constructions, we randomly search for the orthogonal matrices of order $n\times n$ ( in ${\cal O}_n, n\ge 5$) whose rows have weight $\ge d$ each. \\

\noindent
$\bullet$ In construction (\ref{eq:recursive1}), $b=-1$ and $a$ are fixed and $\lambda_i$ takes nonzero values in $\F_q$. Note that with $a$ such that $a^2\equiv -1 \pmod q$, the set $$\{(\lambda,\lambda.a):\lambda \in \F_q\}=\{(x,y) \in \F_q^2:x^2+y^2\equiv 0 \pmod q\}.$$ 
\noindent
One matrix product LCD code is obtained as follows.\\
$\bullet$
$[C_1,\hdots,C_4]{ A}$ with parameters $[16,4,12]_{{\mathbb F}_{11}}$, where\\
$${ A}=\left(\begin{array}{cccc}
2&0&0&0\\
0&3&0&0\\
0&0&6&0\\
0&0&0&4\\
\end{array}\right)
\left(\begin{array}{cccc}
5&5&6&5\\
5&5&5&6\\
6&5&5&5\\
5&6&5&5\\
\end{array}\right),
C_1=\left(\begin{array}{cccc}
9&5&2&10\\
\end{array}\right),$$
$$C_2=\left(\begin{array}{cccc}
10&8&8&2\\
\end{array}\right),
C_3=\left(\begin{array}{cccc}
4&6&2&0\\
\end{array}\right),
C_4=\left(\begin{array}{cccc}
9&6&10&9\\
\end{array}\right).$$
Since there is no MDS linear code with parameters $[16,4,13]$ over ${\mathbb F}_{11}$, this almost MDS code is optimal and has new parameters ever.
\begin{table}\centering{\caption{Optimal and best known LCD codes from random sampling,
*: new code, (d): minimum distance of an existing code}\label{table:code}}
$$
\begin{array}{|c|c|c|c|}
\hline
\text{Over } {\mathbb F}_4&\text{Over } {\mathbb F}_7&\text{Over } {\mathbb F}_{11}&\text{Over } {\mathbb F}_{25}\\
\hline

[4,2,3]_{{\mathbb F}_4}^*&[4,2,3]_{{\mathbb F}_7}^*&[4,2,3]_{{\mathbb F}_{11}}^{*}&[4,2,3]_{{\mathbb F}_{25}}^{*}\\



[5,3,3]_{{\mathbb F}_4}^*&[5,3,3]_{{\mathbb F}_7}^*&[5,3,3]_{{\mathbb F}_{11}}^{*}&[5,3,3]_{{\mathbb F}_{25}}^{*}\\


[6,2,4]_{{\mathbb F}_4}^*&[6,2,5]_{{\mathbb F}_7}^*&[6,2,5]_{{\mathbb F}_{11}}^{*}&[6,2,5]_{{\mathbb F}_{25}}^{*}\\

[6,3,4]_{{\mathbb F}_4}^*&[6,3,4]_{{\mathbb F}_7}^*&[6,3,4]_{{\mathbb F}_{11}}^{*}&[6,3,4]_{{\mathbb F}_{25}}^{*}\\



[7,2,5]_{{\mathbb F}_4}^*&[7,2,6]_{{\mathbb F}_7}^*&[7,2,6]_{{\mathbb F}_{11}}^{*}&[7,2,6]_{{\mathbb F}_{25}}^{*}\\

[7,3,4]_{{\mathbb F}_4}^*&[7,3,5]_{{\mathbb F}_7}^*&[7,3,5]_{{\mathbb F}_{11}}^{*}&[7,3,5]_{{\mathbb F}_{25}}^{*}\\

[7,4,3]_{{\mathbb F}_4}^*&[7,4,4]_{{\mathbb F}_7}^*&[7,4,4]_{{\mathbb F}_{11}}^{*}&[7,4,4]_{{\mathbb F}_{25}}^{*}\\

[7,5,2]_{{\mathbb F}_4}^*&[7,5,3]_{{\mathbb F}_7}^*&[7,5,3]_{{\mathbb F}_{11}}^{*}&[7,5,3]_{{\mathbb F}_{25}}^{*}\\


[8,2,6]_{{\mathbb F}_4}^*&[8,2,7]_{{\mathbb F}_7}^*&[8,2,7]_{{\mathbb F}_{11}}^{*}&[8,2,7]_{{\mathbb F}_{25}}^{*}\\

[8,3,5]_{{\mathbb F}_4}^*&[8,3,6]_{{\mathbb F}_7}^*&[8,3,6]_{{\mathbb F}_{11}}^{*}&[8,3,6]_{{\mathbb F}_{25}}^{*}\\

[8,4,4]_{{\mathbb F}_4}^*&[8,4,5]_{{\mathbb F}_7}^*&[8,4,5]_{{\mathbb F}_{11}}^{*}&[8,4,5]_{{\mathbb F}_{25}}^{*}\\

[8,5,3]_{{\mathbb F}_4}^*&[8,5,4]_{{\mathbb F}_7}^*&[8,5,4]_{{\mathbb F}_{11}}^{*}&[8,5,4]_{{\mathbb F}_{25}}^{*}\\

[8,6,2]_{{\mathbb F}_4}^*&[8,6,3]_{{\mathbb F}_7}^*&[8,6,3]_{{\mathbb F}_{11}}^{*}&[8,6,3]_{{\mathbb F}_{25}}^{*}\\


[9,2,7]_{{\mathbb F}_4}^*&[9,2,7]_{{\mathbb F}_7}^*&[9,2,8]_{{\mathbb F}_{11}}^{*}&[9,2,8]_{{\mathbb F}_{25}}^{*}\\

[9,3,6]_{{\mathbb F}_4}^*&[9,3,6]_{{\mathbb F}_7}^*&[9,3,7]_{{\mathbb F}_{11}}^{*}&[9,3,7]_{{\mathbb F}_{25}}^{*}\\

[9,4,5]_{{\mathbb F}_4}^*&[9,4,5]_{{\mathbb F}_7}^*&[9,4,\ge 5]_{{\mathbb F}_{11}}(6)&[9,4,6]_{{\mathbb F}_{25}}^{*}\\

[9,5,4]_{{\mathbb F}_4}^*&[9,5,4]_{{\mathbb F}_7}^*&[9,5,\ge 4]_{{\mathbb F}_{11}}(5)&[9,5,5]_{{\mathbb F}_{25}}^{*}\\

[9,6,3]_{{\mathbb F}_4}^*&[9,6,3]_{{\mathbb F}_7}^*&[9,6,\ge 3]_{{\mathbb F}_{11}}(4)&[9,6,4]_{{\mathbb F}_{25}}^{*}\\

[9,7,2]_{{\mathbb F}_4}^*&[9,7,2]_{{\mathbb F}_7}^*&[9,7,3]_{{\mathbb F}_{11}}^{*}&[9,7,3]_{{\mathbb F}_{25}}^{*}\\


[10,2,8]_{{\mathbb F}_4}^*&[10,2,8]_{{\mathbb F}_7}^*&[10,2,9]_{{\mathbb F}_{11}}^{*}&[10,2,9]_{{\mathbb F}_{25}}^{*}\\

[10,3,6]_{{\mathbb F}_4}^*&[10,3,7]_{{\mathbb F}_7}^*&[10,3,\ge 7]_{{\mathbb F}_{11}}(8)&[10,3,8]_{{\mathbb F}_{25}}^{*}\\

[10,4,6]_{{\mathbb F}_4}^*&[10,4,6]_{{\mathbb F}_7}^*&[10,4,\ge 6]_{{\mathbb F}_{11}}(7)&[10,4,7]_{{\mathbb F}_{25}}^{*}\\

[10,5,5]_{{\mathbb F}_4}^*&[10,5,5]_{{\mathbb F}_7}^*&[10,5,\ge 5]_{{\mathbb F}_{11}}(6)&[10,5,6]_{{\mathbb F}_{25}}^{*}\\

[10,6,4]_{{\mathbb F}_4}^*&[10,6,4]_{{\mathbb F}_7}^*&[10,6,\ge 4]_{{\mathbb F}_{11}}(5)&[10,6,5]_{{\mathbb F}_{25}}^{*}\\

[10,7,3]_{{\mathbb F}_4}^*&[10,7,3]_{{\mathbb F}_7}^*&[10,7,\ge 3]_{{\mathbb F}_{11}}(4)&[10,7,4]_{{\mathbb F}_{25}}^{*}\\

[10,8,2]_{{\mathbb F}_4}^*&[10,8,2]_{{\mathbb F}_7}^*&[10,8,3]_{{\mathbb F}_{11}}^{*}&[10,8,3]_{{\mathbb F}_{25}}^{*}\\


\hline
\end{array}
$$
\end{table}

\begin{table}\centering{\caption{Optimal and best known LCD codes from matrix product codes,
*: new code, **: new parameters, (d): minimum distance of an existing code}\label{table:code}}
$$
\begin{array}{|c|c|c|c|}
\hline
[C_1,\hdots,C_l]A&\text{Over } {\mathbb F}_q&[C_1,\hdots,C_l]A&\text{Over } {\mathbb F}_q\\
\hline
[C_1,\hdots,C_5]A&[25, 20, 4]_{{\mathbb F}_5}^*&[C_1,\hdots,C_4]A&[16,4,12]_{{\mathbb F}_{11}}^{**}\\

[C_1,\hdots,C_5]A&[25, 15, 7]_{{\mathbb F}_7}^*&[C_1,\hdots,C_5]A&[30,20,\ge 7]_{{\mathbb F}_{11}}^{*}\\

[C_1,\hdots,C_6]A&[30, 18, 8]_{{\mathbb F}_8}^*&[C_1,\hdots,C_5]A&[30,25,\ge 4]_{{\mathbb F}_{11}}^{*}\\

[C_1,\hdots,C_5]A&[25, 10, 11]_{{\mathbb F}_8}^*&[C_1,\hdots,C_6]A&[32,6,\ge 21]_{{\mathbb F}_{11}}^*\\

[C_1,\hdots,C_7]A&[35, 14,\ge 14]_{{\mathbb F}_8}(15)&[C_1,\hdots,C_4]A&[32,16,\ge 11 ]_{{\mathbb F}_{11}}^*\\

[C_1,\hdots,C_5]A&[35, 30, 4]_{{\mathbb F}_9}^*&[C_1,\hdots,C_7]A&[49,7,\ge 34]_{{\mathbb F}_{11}}^*\\
\hline
\end{array}
$$
\end{table}

\begin{table}\centering{\caption{Optimal and best known LCD codes from from projection over a self-dual basis,
*: new code, (d): minimum distance of an existing code}\label{table:code}}
$$
\begin{array}{|c|c|c|c|}
\hline
\text{Over } {\mathbb F}_4&\text{Over } {\mathbb F}_2&\text{Over } {\mathbb F}_{8}&\text{Over } {\mathbb F}_{2}\\
\hline
[4,1,4]_{{\mathbb F}_4}^*&[8,2,5]_{{\mathbb F}_2}^*&[4,1,4]_{{\mathbb F}_{8}}^{*}&[12,3,6]_{{\mathbb F}_{2}}^{*}\\

[4,2,3]_{{\mathbb F}_4}^*&[8,4,\ge 3]_{{\mathbb F}_2}(4)&[4,2,3]_{{\mathbb F}_{8}}^{*}&[12,6,4]_{{\mathbb F}_{2}}^{*}\\

[5,1,5]_{{\mathbb F}_4}^*&[10,2,6]_{{\mathbb F}_2}^*&[5,1,5]_{{\mathbb F}_{8}}^{*}&[15,3,\ge 7]_{{\mathbb F}_{2}}(8)\\

[5,2,3]_{{\mathbb F}_4}^*&[10,4,4]_{{\mathbb F}_2}^*&[5,2,4]_{{\mathbb F}_{8}}^{*}&[15,6,\ge 5]_{{\mathbb F}_{2}}(6)\\

[5,3,3]_{{\mathbb F}_4}^*&[10,6,3]_{{\mathbb F}_2}^*&[5,3,3]_{{\mathbb F}_{8}}^{*}&[15,9,\ge 3 ]_{{\mathbb F}_{2}}(4)\\


[11,2,8]_{{\mathbb F}_4}^*&[22,4,\ge 10]_{{\mathbb F}_2}(11)&[6,2,5]_{{\mathbb F}_{8}}^{*}&[18,6,7]_{{\mathbb F}_{2}}^{*}\\

[11,3,7]_{{\mathbb F}_4}^*&[22,6,9]_{{\mathbb F}_2}^*&[6,3,4]_{{\mathbb F}_{8}}^{*}&[18,9,\ge 5]_{{\mathbb F}_{2}}(6)\\

[11,4,6]_{{\mathbb F}_4}^*&[22,8,\ge 7]_{{\mathbb F}_2}(8)&[6,4,3]_{{\mathbb F}_{8}}^{*}&[18,12,\ge 3]_{{\mathbb F}_{2}}(4)\\

[11,7,4]_{{\mathbb F}_4}^*&[22,14,4]_{{\mathbb F}_2}^*&[7,1,7]_{{\mathbb F}_{8}}^{*}&[21,3, \ge 11]_{{\mathbb F}_{2}}(12)\\

[11,9,2]_{{\mathbb F}_4}^*&[22,18,2]_{{\mathbb F}_2}^*&[7,2,6]_{{\mathbb F}_{8}}^{*}&[21,6,8]_{{\mathbb F}_{2}}^{*}\\

[12,2,9]_{{\mathbb F}_4}^*&[24,4,\ge 11]_{{\mathbb F}_2}(12)&[7,4,4]_{{\mathbb F}_{8}}^{*}&[21,12,\ge 4]_{{\mathbb F}_{2}}(5)\\

[12,3,8]_{{\mathbb F}_4}^*&[24,6,\ge 9]_{{\mathbb F}_2}(10)&[7,5,3]_{{\mathbb F}_{8}}^{*}&[21,15,\ge 3]_{{\mathbb F}_{2}}(4)\\

[12,4,7]_{{\mathbb F}_4}^*&[24,8,8]_{{\mathbb F}_2}^*&[8,1,8]_{{\mathbb F}_{8}}^{*}&[24,3, 13]_{{\mathbb F}_{2}}^{*}\\

[12,8,4]_{{\mathbb F}_4}^*&[24,16,4]_{{\mathbb F}_2}^*&[8,2,7]_{{\mathbb F}_{8}}^{*}&[24,6,\ge 9]_{{\mathbb F}_{2}}(10)\\

[12,9,2]_{{\mathbb F}_4}&[24,18,\ge 3]_{{\mathbb F}_2}(4)&[8,5,4]_{{\mathbb F}_{8}}^{*}&[24,15,4]_{{\mathbb F}_{2}}^{*}\\

\hline
\hline
\text{Over } {\mathbb F}_{16}&\text{Over } {\mathbb F}_2&\text{Over } {\mathbb F}_{32}&\text{Over } {\mathbb F}_{2}\\
\hline
[4,1,4]_{{\mathbb F}_{16}}^*&[16,4,\ge 7]_{{\mathbb F}_2}(8)&[4,1,4]_{{\mathbb F}_{32}}^{*}&[20,5,\ge 8]_{{\mathbb F}_{2}}(9)\\

[4,2,3]_{{\mathbb F}_{16}}^*&[16,8,\ge 4]_{{\mathbb F}_2}(5)&[4,2,3]_{{\mathbb F}_{32}}^{*}&[20,10,\ge 5]_{{\mathbb F}_{2}}(6)\\

[5,1,5]_{{\mathbb F}_{16}}^*&[20,4,\ge 9]_{{\mathbb F}_2}(10)&[5,1,5]_{{\mathbb F}_{32}}^{*}&[25,5,\ge 11]_{{\mathbb F}_{2}}(12)\\

[5,2,4]_{{\mathbb F}_{16}}^*&[20,8,\ge 6]_{{\mathbb F}_2}(7)&[5,2,4]_{{\mathbb F}_{32}}^{*}&[25,10,\ge 7]_{{\mathbb F}_{2}}(8)\\

[5,3,3]_{{\mathbb F}_{16}}^*&[20,12,4]_{{\mathbb F}_2}^*&[5,3,3]_{{\mathbb F}_{32}}^{*}&[25,15,\ge 4 ]_{{\mathbb F}_{2}}(5)\\

\hline
\end{array}
$$
\end{table}
\begin{table}\centering{\caption{Optimal and best known LCD codes from projection over a self-dual basis,
*: new code, (d): minimum distance of an existing code }\label{table:code}}
$$
\begin{array}{|c|c|c|c|}
\hline
\text{Over } {\mathbb F}_{27}&\text{Over } {\mathbb F}_3&\text{Over } {\mathbb F}_{2^m}&\text{Over } {\mathbb F}_2\\
\hline

[4,1,4]_{{\mathbb F}_{27}}^*&[12,3,\ge 7]_{{\mathbb F}_3}(8)&[3,1,3]_{{\mathbb F}_{2^8}}^*&[24,8, 8]_{{\mathbb F}_2}^*\\

[4,2,3]_{{\mathbb F}_{27}}^*&[12,6,\ge 4]_{{\mathbb F}_3}(6)&[4,1,4]_{{\mathbb F}_{2^6}}^*&[24,6,\ge 9]_{{\mathbb F}_2}(11)\\

[5,1,5]_{{\mathbb F}_{27}}&[15,3,9]_{{\mathbb F}_3}^*&[4,1,4]_{{\mathbb F}_{2^8}}^*&[32,8,\ge 11]_{{\mathbb F}_2}(13)\\

[5,2,4]_{{\mathbb F}_{27}}&[15,6,\ge 6]_{{\mathbb F}_3}(7)&[4,2,3]_{{\mathbb F}_{2^8}}^*&[32,16,\ge 8]_{{\mathbb F}_2}(10)\\

[5,3,3]_{{\mathbb F}_{27}}^*&[15,9,4]_{{\mathbb F}_3}^*&[4,2,3]_{{\mathbb F}_{2^9}}^*&[36,18,\ge 6]_{{\mathbb F}_2}(8)\\

[5,4,2]_{{\mathbb F}_{27}}^*&[15,12,2]_{{\mathbb F}_3}^*&[5,1,5]_{{\mathbb F}_{2^7}}^*&[30,6,\ge 12]_{{\mathbb F}_2}(14)\\

[6,1,6]_{{\mathbb F}_{27}}^*&[18,3,\ge 11]_{{\mathbb F}_3}(12)&[5,1,5]_{{\mathbb F}_{2^8}}^*&[40,8,\ge 14]_{{\mathbb F}_2}(16)\\

[6,2,5]_{{\mathbb F}_{27}}^*&[18,6,\ge 8]_{{\mathbb F}_3}(9)&[5,2,4]_{{\mathbb F}_{2^7}}^*&[30,12,\ge 7]_{{\mathbb F}_2}(9)\\

[6,3,4]_{{\mathbb F}_{27}}^*&[18,9,6]_{{\mathbb F}_3}^*&[5,3,3]_{{\mathbb F}_{2^7}}^*&[35,21,\ge 5]_{{\mathbb F}_2}(7)\\

[6,4,3]_{{\mathbb F}_{27}}^*&[18,12,4]_{{\mathbb F}_3}^*&[5,3,3]_{{\mathbb F}_{2^8}}^*&[40,24,\ge 5]_{{\mathbb F}_2}(7)\\

[6,5,2]_{{\mathbb F}_{27}}^*&[18,15,2]_{{\mathbb F}_3}^*&[5,4,2]_{{\mathbb F}_{2^7}}^*&[35,28,\ge 3]_{{\mathbb F}_2}(4)\\

[7,1,7]_{{\mathbb F}_{27}}^*&[21,3,\ge 13]_{{\mathbb F}_3}(14)&[5,4,2]_{{\mathbb F}_{2^8}}^*&[40,32\ge 3]_{{\mathbb F}_2}(4)\\

[7,2,6]_{{\mathbb F}_{27}}^*&[21,6,\ge 10]_{{\mathbb F}_3}(11)&[5,4,2]_{{\mathbb F}_{2^9}}^*&[54,36,\ge 3]_{{\mathbb F}_2}(4)\\

[7,3,5]_{{\mathbb F}_{27}}^*&[21,9,\ge 7]_{{\mathbb F}_3}(9)&[5,4,2]_{{\mathbb F}_{2^{10}}}^*&[50,40,\ge 3]_{{\mathbb F}_2}(4)\\

[7,5,3]_{{\mathbb F}_{27}}^*&[21,15,4]_{{\mathbb F}_3}^*&[6,5,2]_{{\mathbb F}_{2^7}}^*&[42,35,\ge 3]_{{\mathbb F}_2}(4)\\

[8,1,8]_{{\mathbb F}_{27}}^*&[24,3,\ge 15]_{{\mathbb F}_3}(16)&[6,5,2]_{{\mathbb F}_{2^8}}^*&[48,40,\ge 3]_{{\mathbb F}_2}(4)\\

[8,3,6]_{{\mathbb F}_{27}}^*&[24,9,\ge 9]_{{\mathbb F}_3}(10)&[6,5,2]_{{\mathbb F}_{2^9}}^*&[54,45,\ge 3]_{{\mathbb F}_2}(4)\\

[8,5,4]_{{\mathbb F}_{27}}^*&[24,15,\ge 5]_{{\mathbb F}_3}(6)&[6,5,2]_{{\mathbb F}_{2^{10}}}^*&[60,50,\ge 3]_{{\mathbb F}_2}(4)\\

[8,6,3]_{{\mathbb F}_{27}}^*&[24,18,\ge 3]_{{\mathbb F}_3}(4)&[6,5,2]_{{\mathbb F}_{2^{12}}}^*&[72,60,\ge 3]_{{\mathbb F}_2}(4)\\

[9,1,9]_{{\mathbb F}_{27}}^*&[27,3,\ge 17]_{{\mathbb F}_3}(18)&[7,3,5]_{{\mathbb F}_{2^8}}^*&[56,24,\ge 10]_{{\mathbb F}_2}(12)\\

[9,4,6]_{{\mathbb F}_{27}}^*&[27,12,\ge 8]_{{\mathbb F}_3}(9)&[7,6,2]_{{\mathbb F}_{2^8}}^*&[56,48,\ge 3]_{{\mathbb F}_2}(4)\\

[9,5,5]_{{\mathbb F}_{27}}^*&[27,15,\ge 6]_{{\mathbb F}_3}(7)&[7,6,2]_{{\mathbb F}_{2^9}}^*&[63,54,\ge 3]_{{\mathbb F}_2}(4)\\

[9,6,4]_{{\mathbb F}_{27}}^*&[27,18,\ge 5]_{{\mathbb F}_3}(6)&[7,6,2]_{{\mathbb F}_{2^{10}}}^*&[70,60,\ge 3]_{{\mathbb F}_2}(4)\\

[9,7,3]_{{\mathbb F}_{27}}^*&[27,21,\ge 3]_{{\mathbb F}_3}(4)&[7,6,2]_{{\mathbb F}_{2^{12}}}^*&[84,72,\ge 3]_{{\mathbb F}_2}(4)\\

\hline
\end{array}
$$
\end{table}
\section{Conclusion}
In this article we have constructed optimal LCD codes over large finite fields from optimal self-orthogonal codes and from orthogonal matrices.
The latter constructions rely on the presentation of the orthogonal group by generators and relations.  Optimal LCD codes from (random) sampling elements in the orthogonal group  perform better for small lengths which give us more efficient ways for constructing good long LCD codes by matrix product codes as well as by projection over a self-dual basis. 


\begin{thebibliography}{99}


\bibitem{AG} C. Aguilar-Melchor and P. Gaborit,
``On the classification of extremal $[36, 18, 8]$ binary self-dual codes,'' {\em IEEE Transactions on Information Theory}, vol. 54, no 10, pp. 4743--4750, 2008.

\bibitem{BN} T. Blackmore, G.H. Norton, ``Matrix-product codes over $\mathbb F_q$," Appl. Algebra Eng. Commun. Comput.,12 (2001) 477--500.
\bibitem{mag} W. Bosma and J. Cannon, {\em   Handbook of Magma
Functions}, Sydney, 1995.
\bibitem{CheLiu} B. Chen and H. Liu, ``New constructions of MDS codes with complementary duals," https://arxiv.org/abs/1702.07831
\bibitem{CarGui} C. Carlet and S. Guilley, ``Complementary dual codes for counter-measures to side-channel attacks", \emph{Proceedings of the 4th ICMCTA Meeting, Palmela, Portugal}, 2014.

\bibitem{CMCT} C. Carlet, S. Mesnager, C. Tang and Y. Qi,``Euclidean and Hermitian LCD MDS codes," https://arxiv.org/abs/1702.08033v2
 \bibitem{DKOSS} S. T. Dougherty, J-L. Kim, B. Ozkaya , L. Sok and  P. Sole,`` The combinatorics of LCD codes: Linear Programming bound and orthogonal matrices," International Journal of Information and Coding Theory, to appear
\bibitem{GraGul} M. Grassl and T. A. Gulliver, ``On Self-Dual MDS Codes" {\em ISIT 2008}, Toronto, Canada, July 6 --11, 2008
\bibitem{GuOzSo} C. Guneri, B. Ozkaya and Sol\'e, ``Quasi-cyclic complementary dual codes," Finite Fields and Their Applications, Vol. 42, pp. 67--80, 2016.
\bibitem{HufPle} W. C. Huffman, V. Pless, Fundamentals of Error-Correcting Codes, Cambridge University Press, Cambridge (2003).
\bibitem{J} G. J. Janusz,
``Parametrization of self-dual codes by orthogonal matrices,''
{\sl Finite Fields Appl.,} Vol.~13, No.~3,(2007) 450--491.
\bibitem{Jin} L. F. Jin, ``Construction of MDS codes with complementary dual,"{\em IEEE Trans. Inform. Theory,} 2016.
\bibitem{JinXin} L. F. Jin and C. P. Xing, New MDS self-dual codes from generalized Reed-Solomon codes, arXiv:1601.04467v1, 2016.
\bibitem{LS} S. Ling, P. Sol\'e, On the algebraic structure of quasi-cyclic, I: finite fields, {\em IEEE Trans. Inf. Theorey,} 47(2001) 2751--2760.
\bibitem{LL} X. Liu and H. Liu, ``Matrix-Product Complementary dual Codes," https://arxiv.org/abs/1604.03774
\bibitem{Mac} F. MacWilliams, ``Orthogonal matrices over finite fields," Amer. Math. Monthly 76 (1969) 152--164.
\bibitem{MacSlo}  F.J. MacWilliams and N. J. A. Sloane.: The theory of error-correcting codes. Elsevier, 1977.
\bibitem{Massey} J.L. Massey, Linear codes with complementary duals, Discrete Mathematics, 106--107, 337--342, 1992.
\bibitem{S} N. Sendrier, ``Linear codes with complementary duals meet the Gilbert-Varshamov bound," Discrete Math., 304(2004) 345--347.
\bibitem{SSS} M. Shi, L. Sok and P. Sol\'e,``Construction of MDS self-dual codes from orthogonal matrices," https://arxiv.org/abs/1610.07736.

\end{thebibliography}
\end{document}